\documentclass{mn2e}

\newcommand\lapp{\mathrel{\rlap{\lower4pt\hbox{\hskip1pt$\sim$}}
    \raise1pt\hbox{$<$}}}
\newcommand\gapp{\mathrel{\rlap{\lower4pt\hbox{\hskip1pt$\sim$}}
    \raise1pt\hbox{$>$}}}
\newcommand\eapp{\mathrel{\rlap{\raise2pt\hbox{\hskip0pt$\sim$}}
    \lower1pt\hbox{$-$}}}

\begin{document}
\journal{}
\title[Dynamics of perfect fluid Unified Dark Energy models]{Dynamics of perfect fluid Unified Dark Energy models}
\author[Lu{\' \i}s M. G. Be{\c c}a and Pedro P. Avelino]
{Lu{\' \i}s M. G. Be{\c c}a$^{1}$ and  Pedro P. Avelino$^{1,2}$\\ 
$^{1}$Centro de F\'{\i}sica do Porto e Departamento de F\'{\i}sica
da Faculdade de Ci\^encias da Universidade do Porto,\\ \hspace*{1cm}  
Rua do Campo Alegre 687, 4169-007, Porto, Portugal\\
$^{2}$JILA, University of Colorado, Boulder, CO 80309-0440, USA}
\maketitle
\begin{abstract}
In this paper we show that a \emph{one-to-one} correspondence exists 
between 
any dark energy model and an equivalent (from a cosmological point of view, 
in the absence of perturbations) quartessence model in which dark matter 
and dark energy are described by a single perfect fluid. We further show 
that if the density fluctuations 
are small, the evolution of the sound speed squared, $c_s^2$, 
is fully coupled to the evolution of the scale factor and that the 
transition from the dark matter to the dark energy dominated epoch is 
faster (slower) 
than in a standard $\Lambda$CDM model if $c_s^2 > 0$ ($c_s^2 < 0$). 
In particular, we show that the mapping of the simplest quintessence scenario 
with constant $w_Q \equiv p_Q/ \rho_Q$ into a unified dark energy 
model requires $c_s^2 < 0$ at the present time (if $w_Q > -1$) 
contrasting to the Chaplygin gas scenario where one has $c_s^2 > 0$. 
However, we show that non-linear effects severely complicate 
the analysis, in particular rendering linear results invalid even on large 
cosmological scales. Although a detailed analysis of non-linear effects 
requires solving the full Einstein field equations, some  
general properties can be understood in simple terms. In particular, we 
find that in the context of Chaplygin gas models the transition from 
the dark matter to the dark energy  dominated era may be anticipated with 
respect to linear expectations leading to a background evolution similar 
to that of standard $\Lambda$CDM models. On the other hand, in models with 
$c_s^2 > 0$ the expected transition from the decelerating to the 
accelerating phase may never happen.
\end{abstract}

\section{Introduction}

A considerable effort has been devoted, in the past few years, to shape Unified Dark Energy (aka Quartessence) into a coherent phenomenological model. At the moment we are still lacking a solid theoretical motivation in favour of this unified scenario. Yet, the potential outcome of such models is enough to keep a significant amount of papers cyclicly emerging. The drive behind UDE is simple enough to be put into a few words: can dark matter and dark energy share a common unifying nature? Could they be different aspects of the same thing? Either way, an answer to these questions would tell us something of a fundamental nature. 

Evidence for the existence of a dark energy component has been steadily piling over the years. It is clear from observations that most of the matter in the Universe is in a dark non-baryonic form (see, for instance,  Tonry et al.~2003; Riess et al.~2004; Tegmark~2004; Spergel et al.~2006). However, there is at present no direct detection of non-baryonic dark matter or dark energy, their existence merely inferred from their cosmological implications through gravitational effects. Hence, one should not discard the possibility of a single component simultaneously accounting for both dark matter and dark energy. 

Historically, the idea of UDE has sprung from the cosmological properties of the Chaplygin gas (Kamenshchik, Moschella \& Pasquier~2001; Bilic, Tupper \& Viollier~2002; Bento, Bertolami \& Sen~2002), an exotic fluid with an equation of state $p=-A\rho^{-\alpha}$ (where $A$ and $\alpha$ are positive constants). This special fluid has a dual behaviour: it mimics matter ($p \approx 0$) early in the history of the Universe and a cosmological constant much later (with a smooth transition in between) which is highly suggestive of a unified description of dark matter and dark energy.

Although we could assume that the Chaplygin gas is nearly homogeneous in both the radiation and matter eras much like a standard variable $w$-quintessence component (see for example Multamaki, Manera \& Gaztanaga~2003), this would not explore the full potential of the Chaplygin gas as a UDE candidate. Hence, in the absence of any real clues regarding the microscopic properties of quartessence, we will make the simplest possible assumption: \emph{that it can be approximated by a perfect fluid} whose properties are fully specified once the equation of state is known. However, we shall refrain from making any further assumptions regarding the form of its equation of state (other than those which are required by observations that is) and leave the discussion as general as possible (see, however, Bento, Bertolami \& Sen~2004 for a different approach). 

An appealing feature of perfect fluid UDE models is the existence of 
formally equivalent models with a single scalar field, $\phi$, described 
by the action
\begin{equation}
S=\int d^4 x {\sqrt -g} p(X)
\end{equation}
where $p=p(X)$, $\rho=2 X dp/dX-p$ and $u_\mu = \nabla_\mu \phi/{\sqrt {2 X}}$ 
are 
respectively the pressure, the energy density and the four velocity of the 
perfect fluid and $X= \nabla_\mu  \phi \nabla^\mu \phi /2$. If 
$p = X^{(w+1)/(2w)}$ the action describes a perfect fluid with equation 
of state $p = w \rho$ with constant $w \neq 0, -1$ (for $w=1$ one obtains 
the usual massless scalar field action) while if 
\begin{equation}
p(X)=-A^{\frac{1}{1+\alpha}} 
\left(1-(2X)^{\frac{1+\alpha}{2\alpha}}\right)^\frac{\alpha}{1+\alpha}
\end{equation}
one gets a {\it generalized} Born-Infeld action describing the dynamics 
of the {\it generalized} Chaplygin gas. Here we are taking 
$\nabla_\mu \phi$ to be timelike.
In all these models 
a single real scalar field accounts for both dark matter and dark energy.

The predictions of UDE models based on the Chaplygin gas have been tested using observational data including high-$z$ supernovae (Avelino et al.~2003a; Bean \& Dor\'e~2003; Be{\c c}a et al.~2003; Colistete \& Fabris~2005; Bento et al.~2004; Zhu~2004; Bertolami et al.~2005), lensing (Makler, Quinet de Oliveira \& Waga~2003; Silva \& Bertolami~2003; Dev, Jain \& Alcaniz~2004), high precision CMB (Carturan \& Finelli~2003; Bean \& Dor\'e~2003; Amendola et al.~2003; Bento, Bertolami \& Sen~2003) and Large Scale Structure (Fabris, Goncalves \& De Souza~2002a; Fabris, Goncalves \& De Souza~2002b; Bean \& Dor\'e~2003; Be{\c c}a et al.~2003; Sandvik et al.~2004). As we'll later see in more detail, the Chaplygin gas attains very high sound speeds at recent times which has a significant negative impact on small scale structure formation. This was first studied by Sandvik et al. (2004) which concluded that in order to obtain the mass power spectra we observe today, the parameter $\alpha$ in the Chaplygin gas model had to be extremely fine tuned around zero ($\alpha=0$ being the $\Lambda$CDM limit). However, it has been shown by Be{\c c}a et al. (2003)  that this problem could be alleviated (but not solved) by adding baryons to the mixture.   

Yet, there is a caveat to most of these results: linear theory has been taken as a good approximation when comparing with cosmological observations. Ordinarily, this is a valid assumption to make; for example in standard $\Lambda$CDM models the non-linear collapse is not expected to have a large impact on the evolution of the average Universe. However, as discussed previously by Avelino et al. (2004), the small scale structure of the Chaplygin gas may influence the background evolution of the Universe, rendering linear theory invalid even on large cosmological scales. Hence, in order to confront perfect fluid UDE models with observations, linear theory may not enough; we may have to solve the full Einstein field equations which is obviously an enormous task. Nevertheless, in this paper we will show that the overall effect of the non-linear terms on the background evolution of the Universe may be understood using simple considerations.

\section{Local Vs. Global Equations of State}

The dynamics of a homogeneous and isotropic Fried\-mann-Robertson-Walker Universe is partially described by
\begin{eqnarray}
\label{eq1}
\frac{\dot a^2 + k}{a^2}=\frac{8 \pi G}{3}\rho \, , \\
\label{eq2}
\frac{\ddot a}{a}=-\frac{4 \pi G}{3}(\rho+3p) \, ,
\end{eqnarray}
where the dot represents a cosmic time derivative, $a$ is the scale factor, $\rho$ the density, $p$ the pressure and $k$ is a constant (since we are mainly interested in the case of a flat Universe,  we take $k=0$ this point onwards). To fully specify the dynamics, one must supply an extra relation between pressure and density, in other words, an equation of state $p=p(\rho)=w(\rho)\rho$. 

Since the cosmological principle is only approximately valid, so is the description given above. It applies only to the \emph{average} Universe.  In (\ref{eq1}) and (\ref{eq2}), $p$ and $\rho$ refer to \emph{average} global measures of pressure and density obtained by smoothing over large enough regions, as opposed to local measures. Thus, it would probably be better to write them as $\langle p \rangle$ and $\langle \rho \rangle$. Note, however, that the extra prescription $p=w\rho$ is a relation between the local pressure and density and not necessarily between the smoothed versions. This is a subtle and important point.

Consider the perturbative decomposition of pressure and density
\begin{eqnarray}
p&=&\langle p \rangle+\delta p + \cdots \, , \\  \nonumber
\rho&=& \langle \rho \rangle+\delta\rho+\cdots \, .
\end{eqnarray}
The common procedure is to assume that 
$\langle p \rangle = w(\langle \rho \rangle) \langle \rho \rangle$. 
However, a priori, there is no reason to suppose that the relation between the average pressure and the average density is the same as the local one (for a related discussion see Ellis \& Buchert (2005). The Chaplygin gas is an example of this. While locally it behaves as $p=-A\rho^{-\alpha}$, globally it does not (unless $\alpha = 0$): 
\begin{equation}
\langle p \rangle = -A \langle \rho^{-\alpha} \rangle \neq 
-A <\rho>^{-\alpha}\, , 
\end{equation}
except if perturbations are linear ($\delta=\delta\rho/\langle p \rangle \ll 1$) in which case we would have
\begin{eqnarray}
\langle p \rangle &=&  -A \langle \rho^{-\alpha} \rangle = - A\langle \rho \rangle^{-\alpha}\langle 1-\alpha\delta \rangle \\ \nonumber
&=& -A\langle \rho \rangle^{-\alpha}\, .
\end{eqnarray}
This discrepancy between local and global equations of state depends on what is happening to the fluid on small scales meaning that non-linearities are deeply involved. We further discuss this in Section $7$. The important point is that this discrepancy complicates matters substantially. For instance, in the case of the Chaplygin gas, most researchers have naively used the local equation of state to relate $\langle p \rangle$ and $\langle \rho \rangle$ which in general is clearly not a valid assumption.

\section{The simplest UDE model}

The simplest unified dark energy model  one can possibly conceive is that of a perfect fluid with a constant negative pressure: $p=-A$ with constant $A>0$. We can also state this by saying that $w=-A\rho^{-1}$. Incidentally, this is the case of the Chaplygin gas with $\alpha=0$. From (\ref{eq1}) and (\ref{eq2}) it is straightforward to show that 
\begin{equation}
\label{eq2a}
{\dot \rho} + 3\frac{\dot a}{a}(\rho+p)=0 \, .
\end{equation}
This has a simple solution if $p$ is constant. In fact, 
we have that $\rho+p \propto a^{-3}$ and (\ref{eq1}) 
can thus be rewritten as
\begin{equation}
\label{eq2c}
{\rm{H}}^2 = {\rm{H}}_0^2(\Omega_m^{0*}a^{-3} + \Omega_\Lambda^{0*}) \, ,
\end{equation}
where ${\rm{H}}=\dot a/a$ is the Hubble parameter and 
$\Omega_\Lambda^{0*}=1-\Omega_m^{0*}=8\pi G p/3{\rm{H}}_0^2$ 
is the equivalent $\Lambda$ energy fraction today. It is 
also easy to show that 
$w_0 \equiv (p/\rho)_0 = - \Omega_\Lambda^{0*}$ today. Here the index `0' indicates that the variables 
are to be evaluated at the present time, $t_0$. 

The most important feature of this model is that it is 
totally equivalent to $\Lambda$CDM to all orders 
(Avelino, Be{\c c}a, de Carvalho 
\& Martins 2003). This fact 
translates the three year WMAP constraint on 
$\Omega_m^0=0.238^{+0.030}_{-0.041}$ (Spergel et al. 2006) into 
$w_0=-\Omega_\Lambda^{0*}=-0.762^{+0.030}_{-0.041}$. We 
note that the evolution of the equation of state 
of this UDE fluid is given by
\begin{equation}
\label{eqstate}
w= \frac{w_0}{(1+w_0)a^{-3}-w_0}\,,
\end{equation}
making it indeed behave as CDM ($w \sim 0$) at 
early times ($a \sim 0$) and as a cosmological constant 
much later ($a \to \infty$).

\section{Sound speed and background dynamics \label{sound}}

The simplest UDE model just described has a null square sound 
speed $c_s^2=0$ but that will cease to be the case in the 
context of more general models. Indeed, it is a simple matter 
to show, using (\ref{eq1}) and (\ref{eq2}) that
\begin{eqnarray}
\label{eq3}
\dot\rho &=&\frac{3}{4 \pi G}{\rm{H}} \frac{d {\rm{H}}}{dt} \, , \\
\label{eq4}
\dot p &=&-\frac{1}{4 \pi G}
\frac{d}{dt}\left[\left(\frac{\ddot a}{a}+\frac{1}{2}{\rm{H}}^2\right)\right] \,,
\end{eqnarray}
implying a sound speed of
\begin{equation}
\label{eq5}
c_s^2 \equiv \frac{d p}{d \rho}=\frac{1}{3{\rm{H}}}
\frac{d}{d{\rm{H}}}\left[{\rm{H}}^2\left(q-\frac{1}{2}\right)\right] \, ,
\end{equation}
where $q \equiv -\ddot a / (a {\rm{H}}^2)$ is the usual deceleration 
parameter. Let us assume that ${\rm{H}}$ is a decreasing function 
of cosmic time. The sign of the sound speed squared $c_s^2$ today will be 
determined by the way $q$ is evolving. If it is evolving 
sufficiently fast (towards negative values) $c_s^2>0$; 
otherwise $c_s^2<0$. On the other hand, the evolution of $q$ 
is linked to how fast the transition from dark matter to dark 
energy occurs for quartessence. If it is steep enough (faster 
than in $\Lambda$CDM, that is), $c_s^2$ will be positive 
(negative otherwise). Therefore, the sign of $c_s^2$ is 
connected to the background dynamics. 

\section{Extending the simplest UDE model}

A straightforward generalization of the simplest UDE model 
leads to $w=-A\rho^{-(1+\alpha)}$ with a constant 
$\alpha>0$, i.e.~the generalized Chaplygin gas. It is a 
simple matter to show that the sound speed square of this 
fluid is given by
\begin{equation}
c_s^2\equiv\frac{d p}{d \rho}=-\alpha w \,
\end{equation}
and is, in fact, positive for $\alpha>0$. This means 
that the transition from dark matter to dark energy is 
fast enough to make $c_s^2>0$. As we mentioned before, 
one interesting characteristic of the Chaplygin gas is 
that it has a non-zero minimal density. No matter how 
much you expand it, $\rho$ will never drop below a 
certain value: $\rho_m= A^{1/(1+\alpha)}$. When $\rho$ reaches 
this value, however, the pressure will be exactly $-\rho_m$, 
that is, $w=-1$; it behaves as a cosmological constant.

\section{One-to-One Map between UDE and Quintessence\label{1to1map}}

Let us now consider the simplest quintessence model with 
$p_Q=w_Q \rho_Q$ with constant $w_Q$ plus 
pressureless CDM and make a \emph{one-to-one} map into a 
unified dark energy model with density $\rho=\rho_Q+\rho_{cdm}$ 
and pressure $p=p_Q+p_{cdm}=p_Q$ (see also Kamenshchik, Moschella \& 
Pasquier~2001; Bertolami, Sen, Sen \& Silva~2004). If the density 
perturbations are small then the sound speed of this fluid is uniform and 
is given by
\begin{eqnarray}
\label{eq7}
c_s^2 \equiv \frac{d p}{d \rho} &=& \frac{d p_Q}{d \rho_Q} \frac{d \rho_Q}{d \rho} \\ \nonumber
&=& w_Q \frac{1}{1+  \left(\kappa / (1 + w_Q)\right) a^{3w_Q}  }  \ ,
\end{eqnarray}
where $\kappa = \rho_{cdm}^0/\rho_{Q}^0 \approx 0.3-0.4$.
The equation of state of this fluid has the following simple form,
\begin{equation}
\rho=\frac{p}{w_Q}+\rho_{cdm}^0
\left(\frac{p}{w_Q \rho_Q^0}\right)^{1/(1+w_Q)} \, .
\end{equation}
If we the dominant energy 
condition ($w \equiv p/\rho \ge -1$) to be valid at all times then 
we must have $w_Q \ge -1$. In this case $c_s^2 < 0$ at all times.

\section{Non-linear effects \label{nonlinear}}
General relativity is a non-linear theory of gravity which means that, 
unlike in the electromagnetic case, the superposition principle is not 
valid.  However, in many situations we may neglect non-linear terms and 
effectively linearize the field equations. In cosmology, this is often 
possible for large enough scales. However, this is in general not the 
case in the context of perfect fluid UDE models.

\subsection{Case with $c_s^2>0$ \label{positive}}
In this section we illustrate in the simplest possible manner how 
non-linearities can affect the evolution of the Universe 
\emph{even on large cosmological scales}, focusing on the particular 
example of the Chaplygin gas. Here, one should bear in mind than the 
Chaplygin gas has a minimum density $\rho_m$ attainable and 
consequently a minimum pressure $p_m=-\rho_m$ (corresponding to a 
maximum in modulus).

Consider a spherical region of radius $R$ with an average density 
$ \langle \rho \rangle$. If the energy density is \emph{uniformly} 
distributed then we clearly have 
$ \langle p \rangle = -A \langle \rho \rangle^{-\alpha}$. However, 
if the density is not uniformly distributed then, in general, this 
will no longer be valid. Take the case in which the region with $R_1 < r < R$ 
has a smaller density than the region with $r < R_1$. In fact, 
let us assume that $\rho(r<R_1)=N \langle  \rho \rangle$, where $N$ 
is some factor higher than $1$, and 
$\rho(R_1<r<R)=\rho_m=A^{1/(1+\alpha)}$ so that $p(R_1<r<R)=p_m=-\rho_m$. 
The sum of the masses inside the two regions divided by the 
entire volume still has to be $\langle \rho \rangle$ by construction 
which implies that $N$ is equal to
\begin{equation}
N=\left(\frac{R}{R_1}\right)^3+\left(1-\left(\frac{R}{R_1}\right)^3\right)
\frac{\rho_m}{\langle \rho \rangle} \, .
\end{equation}
Now, while the average density inside $R$ is still the same as 
in the uniform case, the average pressure $\langle p \rangle$ 
\emph{is not}. It is simple enough to show that  
\begin{equation}
\langle p \rangle  =  \left(\frac{R_1}{R}\right)^3 \left(N \langle \rho \rangle\right)^{-\alpha} - \left(1-\left(\frac{R_1}{R}\right)^3\right)\rho_m  
\sim   - \rho_m\, ,
\end{equation}
where the approximation is valid for large $N$ (small $R_1/R$). 
If $N\gg1$, the real average 
pressure $\langle p \rangle$, is 
considerably larger (in modulus) than 
$A {\langle  \rho \rangle}^{-\alpha}$, the modulus of the average 
pressure of the Chaplygin gas \emph{in the absence of perturbations}. 
Recall that for the Chaplygin gas, the lower the density, the bigger the 
pressure (in modulus). This is  why the pressure in the lower 
density region will dominate the average pressure inside $R$. 

Hence, the non-linear collapse will make $\langle  p \rangle=p_m=-\rho_m$ 
early on, and will thus anticipate and 
slowdown the transition from dark matter to dark energy, 
thus mimicking the background evolution in the simplest UDE model 
with $\alpha=0$ (or, equivalently, the $\Lambda$CDM scenario). The 
small scale-structure of the Chaplygin gas is interfering with 
the evolution of the Universe on large scales.

Of course, this is an oversimplified picture: we have not taken 
into account the dynamical effects of pressure gradients. In 
high density regions the pressure will be significantly smaller 
(in modulus) than average. Still, we need to take into account 
that gravitational collapse will only be effective on a given scale 
$\lambda$ if $\lambda \gapp c_s H^{-1}$. However, note that once 
the perturbations become non-linear 
pressure will not influence significantly the subsequent dynamics.

\subsection{Case with $c_s^2 < 0$ \label{negative}}

Let us recall what a negative $c_s^2$ means; remember that well 
inside the horizon, linear theory describes the evolution of a
perturbation as a wave:
\begin{equation}
\ddot \delta - c_s^2 \nabla^2\delta \simeq 0 \, ,
\end{equation}
so that $\delta_k \propto \exp\left(i(w t - \mathbf{k\cdot r})\right)$ 
where $w^2=c_s^2 k^2$ (although this is a linear result, 
we will still use it to guide us into the non-linear realm). This 
makes the interpretation of the sign of the sound speed $c_s^2$ 
very straightforward. If $c_s^2>0$, $\delta$ will oscillate as 
an acoustical wave, acting against the formation of voids and 
dense regions. On the other hand, if $c_s^2<0$, the opposite 
will happen: density perturbations in collapsing regions and 
voids get amplified.

Let us now apply to the UDE model of Section $6$ the 
same reasoning we did to the non-uniform Chaplygin gas and 
focus on a background of negative $c_s^2<0$ (the relevant case). 
High density regions 
will tend to behave as pressureless matter with $c_s^2 \approx 0$. 
On the other hand, low density regions with $c_s^2<0$ will 
tend to become increasingly emptier. There is, however, a major 
difference to the 
Chaplygin gas case: there is not a positive minimum density underdense 
regions cannot go below (except if $w=-1$) and, consequently, 
$\rho$ can be arbitrarily close to zero. This has the interesting consequence 
that the average pressure may be close to zero at all times so that the 
Universe may never start to accelerate (despite the linear theory prediction), something
which is clearly inconsistent with current observational evidence.

\section{Conclusions}

In this paper we have shown that there is a one-to-one map 
between any dark energy model and a perfect fluid quartessence model. 
Note, however, that in general these models do not share the 
same underlying physics.
Also, although these models are equivalent in the absence of perturbations, 
this will in general not be the case in the presence of density 
fluctuations. In particular, if the perturbations are small then 
the evolution of $c_s^2$ for the quartessence model is fully coupled to the large-scale dynamics 
of the universe. 
However, we have shown that non-linear effects are of capital importance 
and should be included in quantitative treatments of perfect 
fluid UDE. In fact, small scale structure may alter the 
background evolution of the Universe making the global behaviour of 
quartessence quite different from the one predicted by linear theory. 
In particular, we have found that if $c_s^2 > 0$ then the transition 
from the dark matter to the dark energy dominated epoch may be anticipated 
with respect to linear expectations, leading to a background evolution 
similar to that of standard $\Lambda$CDM models. On the other hand, if 
$c_s^2<0$ the transition from the decelerating to the accelerating phase 
may never happen. There is one obvious consequence of these results: 
if future observations turn out to be incompatible with the background 
evolution predicted in the context standard $\Lambda$CDM models then 
that will also constitute a serious challenge to perfect fluid UDE 
scenarios.

\section*{ACKNOWLEDGMENTS}

We are grateful to Carlos Martins and Paulo Maurício for useful discussions 
on this topic. L.M.G. Be{\c c}a is funded by FCT (Portugal) under grant 
SFRH/BD/9302/2002. Additional support came from FCT under contract 
POCTI/FP/FNU/50161/2003.

\bsp

\begin{thebibliography}{}


\bibitem[Amendola et al.~2003]{Amen} Amendola L., Finelli F., 
Burigana C., Carturan D., 2003, JCAP, 0307, 005 
\bibitem[Avelino et al.~2003a]{Avel03a} Avelino P. P., Be{\c c}a L. M. G., 
de Carvalho J. P. M., Martins C. J. A. P., 2003a, Phys. Rev. D, 67, 
023511
\bibitem[Avelino et al.~2003b]{Avel03b} Avelino P. P., Be{\c c}a L. M. G., 
de Carvalho J. P. M., Martins C. J. A. P., 2003b, JCAP, 0309, 002
\bibitem[Avelino et al.~2004]{Avel04} Avelino P. P., Be{\c c}a L. M. G., 
de Carvalho J. P. M., Martins C. J. A. P., Copeland E. J., 2004, 
Phys. Rev. D, 69, 041301
\bibitem[Bean et al.~2003]{Beam} Bean R., Dore O., 
Gaztanaga E., 2004, Phys. Rev. D, 68, 023515
\bibitem[Be{\c c}a et al.~2003]{Beca03}  Be{\c c}a L. M. G., Avelino P. P., 
de Carvalho J. P. M., Martins C. J. A. P., 2003, Phys. Rev. D, 67, 
101301
\bibitem[Bento et al.~2002]{Bent02} Bento M. C., Bertolami O., 
Sen A. A., 2002, Phys. Rev. D, 66, 043507
\bibitem[Bento et al.~2003]{Bent03} Bento M. C., Bertolami O., 
Sen A. A., 2003, Phys Lett. B, 575, 172
\bibitem[Bento et al.~2004b]{Bent04} Bento M. C., Bertolami O., 
Sen A. A., 2004, Phys. Rev. D, 70, 083519
\bibitem[Bento et al.~2005]{Bent05} Bento M. C., Bertolami O., 
Santos N. M. C., Sen A. A., 2005, Phys. Rev. D, 71, 063501
\bibitem[Bertolami et al.~2004]{Bert} Bertolami O., Sen A. A., 
Sen S., Silva P. T., 2004, MNRAS, 353, 329
\bibitem[Bilic et al.~2002]{Bili} Bilic N., Tupper G. B., 
Viollier R. D., 2002, Phys. Lett. B, 535, 17
\bibitem[Carturan et al.~2003]{Cart} Carturan D., Finelli F., 
2003, Phys. Rev. D, 68, 103501 
\bibitem[Colistete et al.~2005]{Coli} Colistete J., Fabris, J. C., 
2005, Class. Quant. Grav., 22, 2813
\bibitem[Dev et al.~2003]{Dev} Dev A., Jain D., Alcaniz J.S., 2004, 
A\&A, 417, 847
\bibitem[Ellis et al.~2005]{Elli} Ellis G. F. R., Buchert T., 2005, 
Phys. Lett. A, 347, 38
\bibitem[Fabris et al.~2002a]{Fabr02a}  Fabris J. C., Goncalves 
S. V. B., De Souza P. E., 2002a, Gen. Rel. Grav., 34, 53
\bibitem[Fabris et al.~2002b]{Fabr02b}  Fabris J. C., Goncalves 
S. V. B., De Souza P. E., 2002b, Gen. Rel. Grav., 34, 2111
\bibitem[Kamenshchik et al.~2001]{Kame} Kamenshchik A., 
Moschella U., Pasquier V., 2001, Phys. Lett. B, 511, 265
\bibitem[Makler et al.~2003]{Makl} Makler M., de Oliveira Q., 
Waga I., 2003, Phys. Rev. D, 68, 123521
\bibitem[Multamaki et al.~2004]{Mult} Multamaki T., Manera M., 
Gaztanaga E., 2004, Phys. Rev. D, 69, 023004
\bibitem[Riess et al.~2004]{Ries} Riess A. G. et al., 2004, 
ApJ, 607, 665
\bibitem[Sandvik et al.~2004]{Sand}  Sandvik H., Tegmark M., 
Zaldarriaga, M., Waga I., Phys. Rev. D, 69, 123524
\bibitem[Spergel et al.~2006]{Sper} Spergel D. N. et al., 2006, 
astro-ph/0603449 
\bibitem[Silva et al.~2003]{Silv} Silva P. T, Bertolami O., 
2003, ApJ, 599, 829
\bibitem[Tegmark et al.~2004]{Tegm} Tegmark M. et al., 2004, 
Phys. Rev. D, 69, 103501
\bibitem[Tonry et al.~2003]{Tonr} Tonry J. L. et al., 2003, 
ApJ, 594, 1
\bibitem[Zhu~2004]{Zhu}  Zhu Z. H., 2004, A\&A, 423, 421
\end{thebibliography}
\end{document}